\documentclass[10pt,twocolumn,letterpaper]{article}

\usepackage{cvpr}              %

\usepackage{suffix}
\usepackage{url}
\usepackage{tikz}
\usepackage{amsmath}
\usepackage{subcaption}
\usepackage{graphicx}
\usepackage{amssymb}
\usepackage{dblfloatfix}
\usepackage{pgfplots}
\usepackage{makecell}
\usetikzlibrary{calc}

\usetikzlibrary{arrows.meta, positioning}

\definecolor{cvprblue}{rgb}{0.21,0.49,0.74}
\usepackage[pagebackref,breaklinks,colorlinks,allcolors=cvprblue]{hyperref}

\def\paperID{7} %
\def\confName{CVPR - PBVS}
\def\confYear{2026}

\title{ACCOR: Attention-Enhanced Complex-Valued Contrastive Learning for Occluded Object Classification Using mmWave Radar IQ Signals}

\author{Stefan Hägele, Adam Misik, Constantin Patsch, Eckehard Steinbach\\
School of Computation, Information and Technology, Chair of Media Technology
\\
Munich Institute of Robotics and Machine Intelligence\\
Technical University of Munich
\\
{\tt\small \{stefan.haegele, adam.misik, constantin.patsch, eckehard.steinbach\}@tum.de}}

\begin{document}
\maketitle
\begin{abstract}
Millimeter-wave (mmWave) radar provides robust sensing under adverse conditions and can penetrate thin materials for non-visual perception in industrial and robotic settings.
Recent work with MIMO mmWave radar has demonstrated its ability to penetrate cardboard packaging for occluded object classification.
However, existing models leave room for improvement and extensions across different sensing frequencies.
Building on recent work with MIMO radar for occluded object classification, we propose ACCOR, an attention-enhanced complex-valued contrastive learning approach for radar, enabling robust occluded object classification.
ACCOR processes complex-valued IQ radar signals via a complex-valued CNN backbone, a multi-head attention layer and a hybrid loss. The hybrid loss combines a weighted cross-entropy term with a supervised contrastive term.
We extend an existing 64 GHz dataset with a new 67 GHz subset and evaluate performance across both bands. ACCOR achieves 96.60 \% accuracy at 64 GHz and 93.59 \% at 67 GHz on 10 objects, surpassing prior radar-specific and adapted image models.
Results demonstrate the benefits of integrating complex-valued deep learning, attention, and contrastive learning for mmWave radar-based occluded object classification.
\\
The dataset can be found at
\url{https://github.com/haegels/radar-IQ-datasets}.
\end{abstract}
    
\section{Introduction}
\label{sec:intro}
In recent years, millimeter-wave (mmWave) radar has gained increasing popularity across various sensing domains. Due to its low cost and wide availability, it provides robust performance in applications such as automotive, tracking, human activity or gesture recognition. Moreover, mmWave radar is often integrated into sensor fusion frameworks to enhance perception, particularly in automotive systems \cite{human, tracking, tracking2, gesture1, gesture2, gesture3, philipp, philipp2}.
Unlike optical sensors such as LiDAR or RGB cameras, mmWave radar operates reliably under adverse environmental conditions including fog, smoke, rain, or complete darkness, where optical methods typically degrade \cite{survey, mengchen, smoke, smoke2}. This enables more robust perception across all application domains where mmWave radar sensors are employed.
Furthermore, operating in the 30–300 GHz range of the non-visible electromagnetic spectrum, it can penetrate lightweight, non-metallic, and weakly reflective materials such as fabric, cardboard, or plastic.
This property provides robust mmWave radar-based perception for occluded object classification, making it well-suited for applications such as automated inspection, concealed object detection, and inventory management in industrial environments. In conveyor belt setups, mmWave radar can identify packaged objects, allowing robots to handle sorting, inspection, and further processing. These are capabilities that conventional optical sensors cannot typically provide.
Existing radar-based concealed object classification methods, however, remain constrained by their reliance on large-scale antenna array imaging scanners, either physical or virtual, which limits scalability for compact industrial automation systems \cite{imaging1, imaging2}. These approaches also typically require manual inspection or image-processing algorithms for object detection and classification.
As a first step towards more practical solutions, previous work showed that occluded object detection and classification can be achieved using a compact and cheap mmWave MIMO radar, direct IQ signal processing (rather than range-Doppler- or range-angle map image processing) and a learning-based classification approach \cite{stefan}.
Using various Convolutional Neural Network (CNN) architectures designed for complex-valued radar signals, prior work successfully classified occluded objects within packaging boxes.
However, the benchmark model's 3D convolutions impose high computational cost, and the dataset is limited to a single frequency band.
Despite recent general progress, IQ signal-based approaches have not been systematically examined across different frequency bands. To address this gap, we introduce ACCOR, a deep learning model featuring a compact complex-valued CNN backbone with integrated self-attention for signal refinement, a hybrid loss function, and new 67 GHz dataset to study frequency-dependent penetrability.
Complex-valued CNNs outperform their real-valued counterparts when processing complex-valued input, even when input adaptations are applied \cite{cplx_cnn, cplx_cnn2}.
Moreover, the network is trained using a hybrid loss that combines a cross-entropy and a supervised contrastive loss to enhance class separability and overall classification performance, which is beneficial given that radar signals are inherently very similar to one another.
The proposed framework aims to improve both efficiency and robustness, enabling compact occluded object classification and advancing compact mmWave radar toward practical deployment in industrial automation and logistics.
The findings of this work can be carried forward to develop a compact automated radar-based perception system for any industrial setup.
\\
Our contributions are as follows:
\begin{itemize}
    \item We design a compact complex-valued CNN backbone with integrated self-attention, tailored to exploit amplitude- and phase information in preprocessed radar IQ signals.
    \item We introduce a hybrid loss function that combines cross-entropy with supervised contrastive learning, improving class separability and overall classification accuracy.
    \item We extend existing benchmarks with a new sub-dataset collected at 67 GHz, enabling a comparative analysis of two different frequency bands for occluded object classification.
\end{itemize}
The remainder of this paper is organized as follows: Section \ref{related_work} reviews related work in mmWave radar. Section \ref{meth} presents the proposed methodology, while Section \ref{results} reports results, model comparisons and ablations. Finally, Section \ref{conclusio} concludes the paper and discusses potential directions for future research.

\section{Related Work}
\label{related_work}

As mentioned, radar, and especially mmWave radar, has gained considerable attention in the last decade.
One reason is the increasing availability of compact, high-precision radar solutions on the market.
The automotive sector remains the largest application domain for mmWave radar, where its robustness under adverse conditions such as rain and fog makes it a valuable complement to camera and LiDAR sensors for reliable perception \cite{philipp, philipp2, survey, fusion1}. While cameras and LiDAR degrade significantly under such adverse conditions, radar can still perceive the environment due to its ability to penetrate lightweight materials or particles, such as rain \cite{weather}.
Furthermore, with Doppler-based processing, mmWave radar is widely used for tracking applications in both indoor and outdoor environments \cite{tracking, tracking2, tracking3}. 
This is especially appealing since radar naturally preserves privacy compared to camera sensors.
Another field of mmWave radar research is human activity understanding and gesture recognition.
Like tracking, this area relies on Doppler- or micro-Doppler processing to gain information about movement and activities of humans \cite{gesture1, gesture2,gesture3, gesture4, gesture5}.
When very high resolution is required, commonly available radar sensors are limited by their small number of transmit and receive antennas. Since a larger number of antennas provides higher spatial resolution, radar imaging is often achieved synthetically by scanning a large area with a moving radar to form a large virtual antenna array over a certain trajectory. These virtual arrays can then be exploited for higher-resolution imaging, as demonstrated by approaches such as Synthetic Aperture Radar (SAR) for earth observation \cite{sar1,sar2}.
However, this approach can also be applied on a smaller scale using testbeds to scan and image smaller objects \cite{imaging3, imaging4}.
A remedy to these high-effort imaging techniques is the use of radars with larger number of transmit- and receive antennas, providing many virtual channels without requiring active device movement \cite{vayyar}.
Such sensor was used in previous data collection and is further utilized and extended in this work \cite{stefan}. Publicly available datasets using these high-resolution sensors remain very limited up to now. This creates the need for further and more diverse data collection.
Despite the aforementioned advances in mmWave radar, most existing approaches rely on preprocessed range–Doppler images, or on CFAR-generated radar point clouds as input for deep learning models. This shifts radar research toward computer vision rather than actual signal processing methods.
However, research directly leveraging down-converted complex-valued IQ radar signals, especially with deep learning, remains limited \cite{smcnet, radarcnn, iq}. This is mainly due to the lack of publicly available datasets with complex-valued radar IQ signals, as opposed to their preprocessed image counterparts, even though these preprocessing steps commonly remove information from the IQ signal. Furthermore, the penetration ability of mmWave radar applies not only to rain and fog but also to light materials such as cardboard or thin drywall \cite{stefan, wall1, wall2}.
Due to its popularity surge, mmWave radar also arrived as very helpful perception assistance in robotics and automation, as extensively surveyed in \cite{tro_radar}. This includes tasks such as localization, mapping, and object classification in various scenarios. Due to its compactness, mmWave radar can be easily deployed on different moving platforms such as robots or drones \cite{robots, drones}.
Beyond traditional supervised learning for radar, recent work has explored contrastive and attention-based models, but their application in mmWave radar remains limited, particularly in IQ signal-based processing.
However, the informational content in IQ signals is higher than in its processed counterparts, suggesting that attention-based and contrastive approaches can extract richer information directly from the radar IQ signal.
Moreover, most contrastive learning research in the radar domain has focused on image-based Synthetic Aperture Radar (SAR) \cite{cont1, cont2}.
The same holds for radar transformer models and, more broadly, attention-based radar approaches \cite{philipp,philipp2,transrad, attention}.
To conclude, radar can provide significant perception advantages for industrial automation and logistics, motivating this work on occluded object classification.
We use the IQ data from \cite{stefan}, complemented by an additional sub-dataset collected at a different frequency band. Together with a complex-valued CNN, this forms the starting point of the work presented in this paper.

\section{Methodology}
\label{meth}
\subsection{Preprocessing and Setup}
\begin{figure}
    \centering
    \begin{tikzpicture}[scale=0.7]
\definecolor{softgreen}{RGB}{120, 180, 140}
\definecolor{softblue}{RGB}{100, 150, 200}
\definecolor{softorange}{RGB}{240, 170, 100}
\definecolor{softred}{RGB}{210, 100, 100}
\definecolor{stronggreen}{RGB}{60, 160, 100}
\definecolor{strongblue}{RGB}{70, 130, 200}
\definecolor{strongorange}{RGB}{230, 140, 60}
\definecolor{strongred}{RGB}{200, 60, 70}
    \foreach \x in {0,1,2,...,19} {
        \fill[stronggreen] (\x*0.25+0.5,0) circle (2pt); %
    }

    \foreach \y in {0,1,2,...,19} {
        \fill[strongred] (5.75,-\y*0.25-0.5) circle (2pt); %
    }

    \foreach \x in {0,1,2,...,19} {
        \foreach \y in {0,1,2,...,19} {
            \fill[strongblue] (\x*0.25+0.5,-\y*0.25-0.5) circle (2pt); %
        }
    }

\fill[stronggreen] (0,-6) circle (2pt);

\node[text=stronggreen] at (1.5,-6) {\footnotesize{Rx antenna}};

\fill[strongred] (3,-6) circle (2pt);
\node[text=strongred] at (4.5,-6) {\footnotesize{Tx antenna}};

\fill[strongblue] (2,-7) circle (2pt);
\node[text=strongblue] at (3.5,-7) {\footnotesize{Virtual channel}};
    
\end{tikzpicture}
    \caption{Virtual channels formed by each antenna pair in the L-shaped $20 \times 20$ antenna array.}
    \label{virtual_array}
\end{figure}
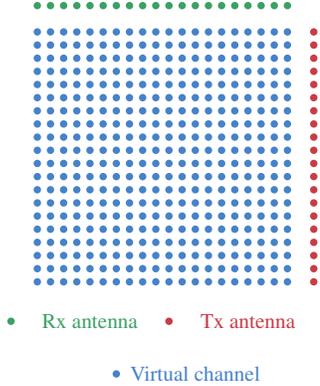

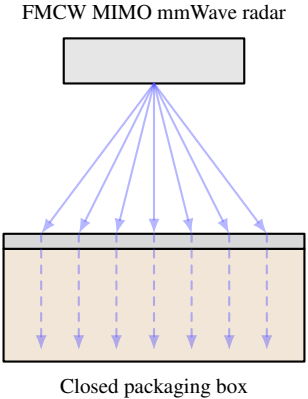
\begin{figure}
    \centering
    \begin{tikzpicture}[scale=1.0, line cap=round, line join=round]

\draw[fill=brown!20, thick] (-2,0) rectangle (2,1.5);     %
\draw[fill=gray!30, thick]  (-2,1.5) rectangle (2,1.7);  %
\node[font=\footnotesize] at (0,-0.35) {Closed packaging box};

\draw[fill=gray!20, thick] (-1.2,3.7) rectangle (1.2,4.3);
\node[font=\footnotesize] at (0,4.65) {FMCW MIMO mmWave radar};

\foreach \x in {-1.5,-1,-0.5,0,0.5,1,1.5}{
  \draw[-{Latex[length=2mm]}, thick, blue!70, opacity=0.4] (0,3.7) -- (\x,1.7);
  \draw[-{Latex[length=2mm]}, thick, blue!70, opacity=0.4, dashed] (\x,1.7) -- (\x,0.15);
}

\end{tikzpicture}
    \caption{Sensing setup: a top-mounted mmWave radar sensing the contents of a closed packaging box.}
    \label{setup}
\end{figure}

\begin{figure}
    \centering
    \includegraphics[width=0.35\linewidth]{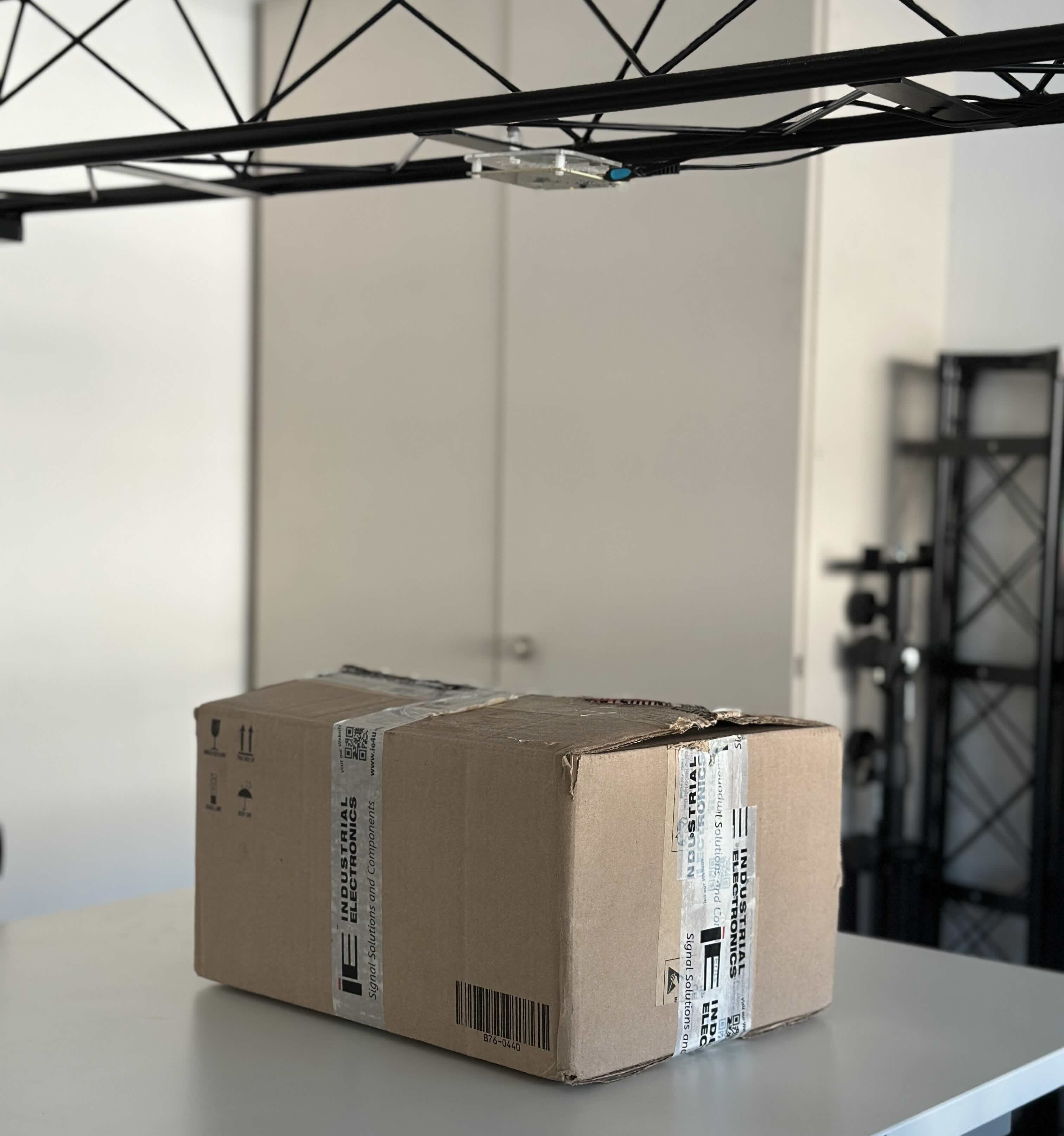}
    \caption{Image of the measurement setup from \cite{stefan}.}
    \label{real-setup}
\end{figure}

The radar sensor used for data collection is a 62-69 GHz frequency-modulated continuous wave (FMCW) MIMO mmWave imaging radar from Vayyar and Mini-Circuits \cite{vayyar}. This radar is equipped with 20 transmit (Tx) and 20 receive (Rx) antennas, yielding 400 virtual channels from a single measurement shot.
Tx and Rx antennas are arranged in an L-shaped position, which generates a large virtual channel array, as depicted in Fig.~\ref{virtual_array}.
The transmit signal of a single channel is an FMCW chirp, and can be expressed by the following formula
\begin{equation}
    s_{c}[t] = a[t] e^{j2\pi f_c[t]t},
\end{equation}
while every virtual channel in Fig.~\ref{virtual_array} (blue) is sampled with $N=100$ points over one sampling period. $a[t]$ denotes the amplitude, $f_c[t]$ the time-dependent frequency, and $c = 0,1,2,\ldots,399$ the channel index.
Hence, a single data sample consists of 400 arrays of complex-valued IQ signals.
The mathematical dimension of a 
data sample is given by
\begin{equation}
    \boldsymbol{s} \in \mathbb{C}^{(Rx \cdot Tx)\times N},
\end{equation}
where $Rx \cdot Tx=400$ and $N=100$.
The IQ signal serves as the fundamental input to the overall model architecture.
To refine the range-specific features of this signal collection, a Fourier transform is applied, generating the complex-valued range profile of the FMCW radar IQ signal. Applied as a Fast Fourier Transform (FFT), the Discrete Fourier Transform (DFT) is given by
\begin{equation}
S_{c}[k] = \sum_{t=0}^{N-1} s_c[t] \, e^{-j 2 \pi \frac{kt}{N}}, \quad c = 0,1,2,\ldots,399,
\end{equation}
with $N=100$.
For deep learning models such as CNNs, the range profile provides richer and more discriminative features than the raw time-domain IQ signal, as shown in \cite{smcnet}.
The schematic setup used for data collection is depicted in Fig.~\ref{setup}, where the distance from sensor to box is around 50 cm.
The real setup can be seen in Fig.~\ref{real-setup}.
The box with the occluded object is placed underneath the mmWave sensor and the sample is taken from this position.
Due to the penetration capability of mmWave radar, the reflected signal carries information about the object inside the box.

\subsection{Model Design}
\subsubsection{CNN and Multi-Head Attention}
\begin{figure*}
    \centering
    \includegraphics[width=\textwidth]{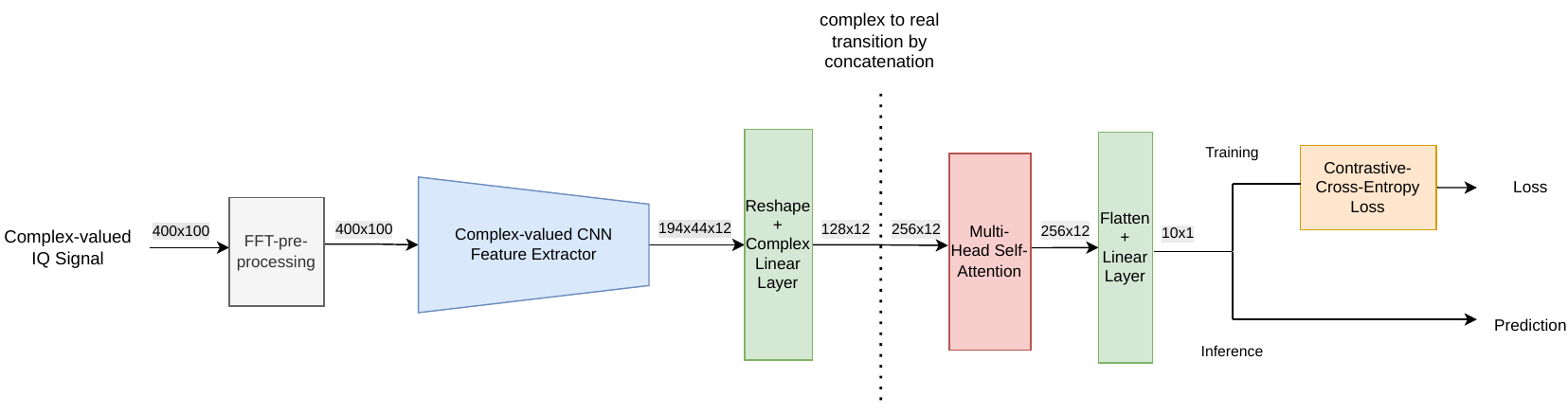}
    \caption{ACCOR model consisting of FFT preprocessing, a CNN backbone, and a multi-head attention layer, trained with the proposed hybrid loss function.}
    \label{model-overview}
\end{figure*}

\begin{figure}
    \centering
    \includegraphics[width=0.5\textwidth]{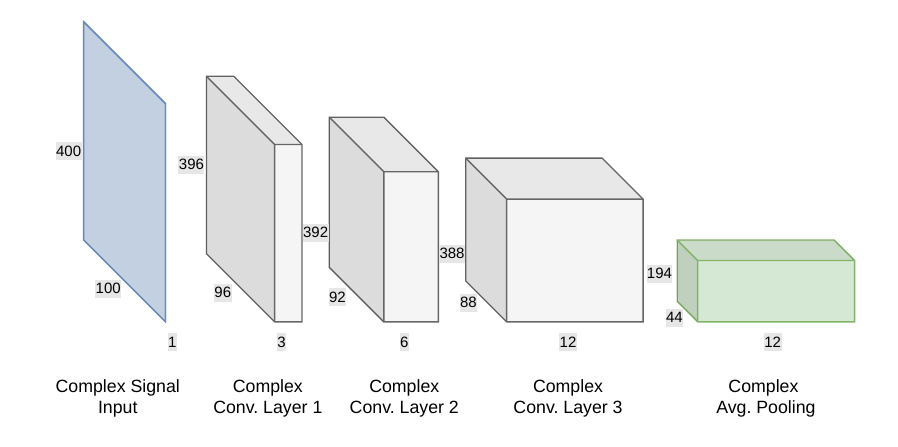}
    \caption{Complex-valued CNN backbone with three layers, kernel size 5, and average pooling, kernel size $2\times2$.}
    \label{cplxcnn-backbone}
\end{figure}

Our model backbone is a complex-valued CNN that processes and operates directly on complex-valued inputs. Since the FFT-transformed IQ signal remains complex-valued, extracting meaningful amplitude and phase features requires complex-valued processing to preserve present signal correspondences lost otherwise with real-valued operations.
Separating the I and Q components into distinct real-valued channels discards the inherent phase relationships and cross-correlations between them, reducing the signal’s feature richness for deep learning models. Real-valued processing also fails to preserve the rotational invariance and magnitude-phase coupling that complex representations naturally capture.
Convolutions, batch normalization, and activation functions are all applied in the complex domain \cite{cplx_cnn}.
For kernel $k = a+jb \in \mathbb{C}$ and input $x = c+jd \in \mathbb{C}$, the convolution is defined by
\begin{equation}
    k \cdot x = c \cdot a - d \cdot b + j(c \cdot b  +d \cdot a).
\end{equation}
The complex-valued batch normalization $cBN$ of $z \in \mathbb{C}$ is expressed as
\begin{equation}
    cBN(z) = BN(Re\{z\}) + j BN(Im\{z\}),
\end{equation}
which preserves the phase information.
The complex-valued ReLU activation function $cReLU$ is defined as
\begin{equation}
    cReLU(z)
    \begin{cases}
    z, &\text{if $Re\{z\}$, $Im\{z\}$} \geq 0\\
    0, &\text{else}.
    \end{cases}
\end{equation}
The single kernels for every complex-valued layer have kernel size $5\times5$.
After feature extraction, the features are embedded into a token vector of length $D=256$. This dimensionality empirically yields the best performance. Before the projection phase, the complex-valued feature vector is transferred into the real domain.
This is achieved by projecting the feature maps into a vector of length 128 using a complex-valued linear layer and concatenating the real and imaginary parts to obtain a length of $D$.
This simplifies subsequent processing while preserving the information extracted from the complex-valued backbone.
The token vector is then fed into a multi-head self-attention layer with 16 heads.
This mechanism is used to refine radar features by enabling the model to capture diverse feature dependencies in range as well as angle domain through multiple parallel attention heads.
Fig.~\ref{model-overview}  presents the overall model architecture, while Fig.~\ref{cplxcnn-backbone} illustrates the complex-valued CNN backbone.
\begin{figure*}[!tbp]
\begin{equation} \tag{9}
\ell_{\kappa}
= \frac{1}{|\{\, i:\,|P(i)|>0 \,\}|}
\sum_{i:\,|P(i)|>0}
\left(
-\frac{1}{|P(i)|}\sum_{j\in P(i)}
\log
\frac{\exp\!\left(\frac{z_i^\top z_j}{\tau}\right)}
{\sum_{k\in A(i)} \exp\!\left(\frac{z_i^\top z_k}{\tau}\right)}
\right)
\label{contr}
\end{equation}
with \(z_i=\frac{f(x_i)}{\lVert f(x_i)\rVert_2}\), \(A(i)=\{k\mid k\neq i\}\), \(P(i)=\{j\in A(i)\mid y_j=y_i\}\), and temperature \(\tau>0\).
\end{figure*}
\subsubsection{Loss Function}
A hybrid loss function is employed for the classification task. While prior work relied on the standard cross-entropy loss, we adopt a hybrid loss combining weighted cross-entropy loss with contrastive loss, enhancing feature separability and robustness.
The contrastive term employs a supervised contrastive approach without data augmentation, contrasting positive and negative samples against a reference anchor (reference sample).
Since preprocessed radar signals are generally very similar to each other, the contrastive component encourages greater separation of different class samples in the feature space, while the cross-entropy component provides information-theoretically sound label predictions.
The total loss is defined as
\begin{equation}
    \ell_{\text{total}} = (1-\alpha)\,\ell_{\chi} + \alpha \, \ell_{\kappa},
    \label{loss}
\end{equation}
where $\ell_{\chi}$ and $\ell_{\kappa}$ denote the cross-entropy and contrastive loss, respectively.
The parameter $\alpha$ serves as the weighting factor between the two loss terms.
Furthermore, the cross-entropy loss $\ell_{\chi}$ is defined in Eq.~\ref{crossentropy} and the contrastive loss $\ell_{\kappa}$ in Eq.~\ref{contr}.
\begin{equation}
    \ell_{\chi} = H(\boldsymbol{\hat{y}},\boldsymbol{y}) = - \sum_{j=1}^{C} y_{j} \log(\hat{y}_{j})
    \label{crossentropy}
\end{equation}
$\boldsymbol{\hat{y}}$ and $\boldsymbol{y}$ are the predicted- and target distributions, respectively, and $C$ is the number of object classes.
For Eq.~\ref{contr}, the parameters are defined as follows:
\begin{itemize}
    \item $z_i$: L2-normalized feature of sample $i$, $z_i = f(x_i) / ||f(x_i)||$
    \item $f(x_i)$: Model output (pre-normalized feature vector) for input $x_i$.
    \item $y_i$: Class label (integer) of sample $i$.
    \item $\tau$: Temperature $>0$.
    \item $A(i)$: All non-self indices $\{k \mid k \ne i\}$; candidates contrasted against anchor $i$.
    \item $P(i)$: Positive indices $\{ j \in A(i) \mid y_j = y_i \}$; same-class samples for anchor $i$.
    \item $|P(i)|$: Number of positives for anchor $i$. 
\end{itemize}

\section{Results and Performance Comparison}
\label{results}
\subsection{Overall Evaluation}
\begin{figure}
    \centering
    \includegraphics[width=\linewidth]{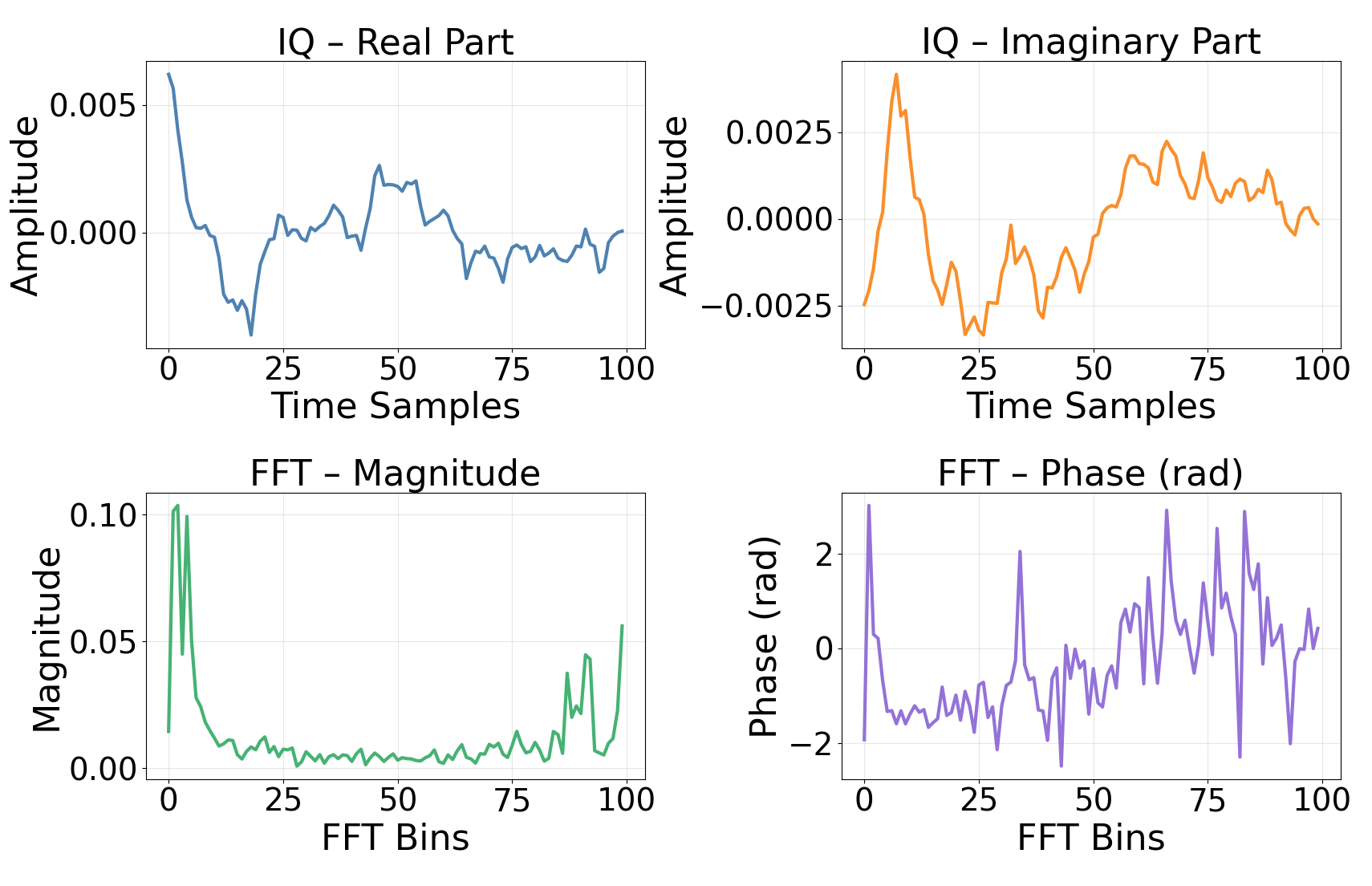}
    \caption{Visualization of the raw IQ signal (top) and its FFT-processed representation (bottom) for a deodorant sample (single channel).}
    \label{fig:example-sample}
\end{figure}

The evaluation of the occluded object classification models is conducted using two subsets of data.
One subset, introduced in \cite{stefan}, consists of samples collected at 64 GHz center frequency. 
We extended the original dataset with an additional subset collected at 67 GHz center frequency and with the same measurement setup including hardware, occlusion types and objects. Both subsets use 4 GHz bandwidth. This extension enables a comparative analysis of the penetration capabilities of the two frequency bands.
The object collection consists of 10 everyday items of similar size, which were placed individually in different orientations inside a packaging box before being scanned by the mmWave radar. Different object orientations are used to introduce structural variability in the data, promoting a more generalizable representation. The testbed itself remains fixed, ensuring consistent background statistics comparable to those in industrial environments.
The box is made of standard cardboard typically used for conventional packaging. It is fully closed and sealed before each measurement is taken by the radar sensor.
\begin{figure}
    \centering
    \includegraphics[width=0.35\linewidth]{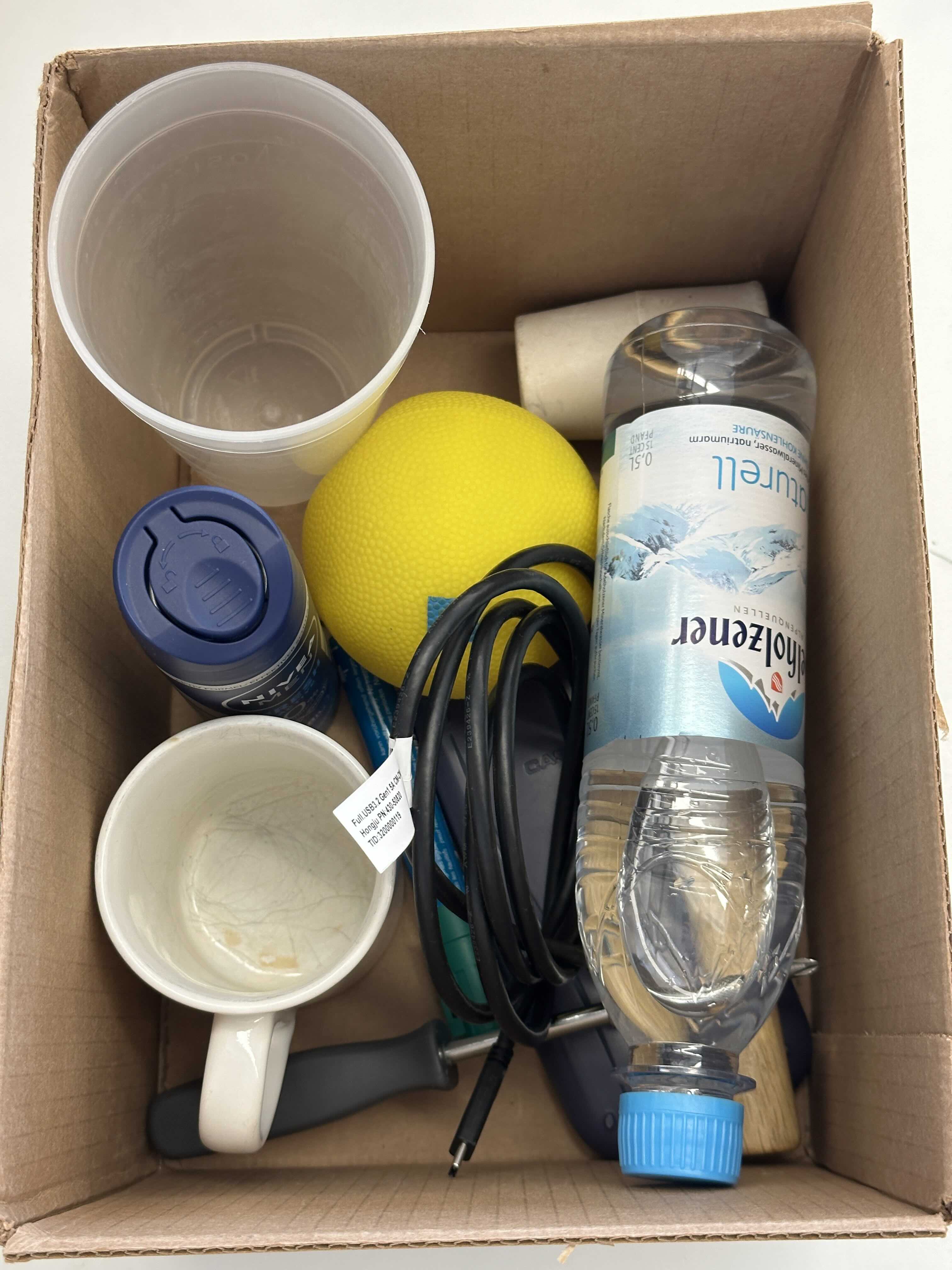}
    \caption{Collection of the 10 objects used in the experiments.}
    \label{objects}
\end{figure}
The object collection contains the items listed in Table~\ref{tab:objects} and the objects are depicted in Fig.~\ref{objects}.
\begin{table}[htbp]
\centering
\caption{Experimental objects (10 classes).}
\label{tab:objects}
\small
\begin{tabular}{ll}
\toprule
Hammer & Mug \\
Screwdriver & Tape roll \\
Deodorant & Water bottle \\
Calculator & Plastic cup \\
Coiled cable & Ball \\
\bottomrule
\label{tab:objects}
\end{tabular}
\end{table}
An example sample of the deodorant is depicted in Fig.~\ref{fig:example-sample}. Next to near-field clutter, the object peak in the bottom left figure is clearly visible on the right side of the graph.
The dataset is divided into training and test sets using the conventional 80/20 ratio. 
Training and testing are conducted separately with the 64 GHz and 67 GHz datasets.
We evaluated over 10 training runs with different random seeds to ensure robustness to initialization with batch size 32 and over 35 epochs each. All reported results are averaged over these runs to improve statistical reliability.
Both datasets comprise 1000 IQ samples total (40,000 sample points per sample), split evenly as 500 per frequency band, with uniform distribution across all 10 classes (50 samples/class/band).
Our empirical evaluation identified $\tau= 0.1$ as the best temperature value, ensuring stable training and reliable convergence. For different values of $\alpha$, $\alpha=0.4$ achieved the best overall performance across both frequency bands.
To demonstrate the effectiveness of our approach, we compare its overall classification accuracy with several other classification models.
The evaluation tables are divided into two categories. Table~\ref{results-1} presents the performance metrics of various classification models capable of directly processing complex-valued radar signal inputs. All baseline models were implemented and trained using the same protocol as ACCOR to ensure fair comparability.
The results show enhanced classification accuracy attained by our proposed model.
Fig.~\ref{fig:conf-matrices} shows confusion matrices for both frequency bands at $\alpha=0.4$. The model struggles most with ball identification at 64 GHz, hammer and plastic cup at 67 GHz. Nonetheless, the model configuration achieves robust per-class classification overall, with several classes perfectly classified.
However, different failure patterns across both frequency bands confirm that 64 GHz and 67 GHz provide different features.
\begin{figure*}
    \centering
    \begin{subfigure}[b]{0.48\linewidth}
        \centering
        \begin{tikzpicture}[scale=0.7]
    \begin{axis}[
            colormap={lightpurple}{color=(white) rgb255=(74,145,199)},
            xlabel=Prediction (in \%),
            xlabel style={yshift=-45pt},
            ylabel=Label,
            ylabel style={yshift=25pt},
            xticklabels={Hammer, Screwdriver, Deodorant, Calculator, Water Bottle, Plastic Cup, Cable, Ball, Mug, Tape},
            xtick={0,...,9},
            xtick style={draw=none},
            yticklabels={Hammer, Screwdriver, Deodorant, Calculator, Water Bottle, Plastic Cup, Cable, Ball, Mug, Tape},
            ytick={0,...,9},
            font =\small,
            ytick style={draw=none},
            enlargelimits=false,
            xticklabel style={
              rotate=90
            },
            nodes near coords={\pgfmathprintnumber\pgfplotspointmeta},
            nodes near coords style={
                yshift=-7pt,
                font=\scriptsize
            },
        ]
        \addplot[
            matrix plot,
            mesh/cols=10,
            point meta=explicit,draw=gray
        ] table [meta=C] {
            x y C
            0 0 100.00
            1 0 0.00
            2 0 0.00
            3 0 0.00
            4 0 0.00
            5 0 0.00
            6 0 0.00
            7 0 0.00
            8 0 0.00
            9 0 0.00
            
            0 1 0.00
            1 1 100.00
            2 1 0.00
            3 1 0.00
            4 1 0.00
            5 1 0.00
            6 1 0.00
            7 1 0.00
            8 1 0.00
            9 1 0.00
            
            0 2 0.00
            1 2 0.00
            2 2 93.94
            3 2 4.04
            4 2 0.00
            5 2 0.00
            6 2 0.00
            7 2 0.00
            8 2 2.02
            9 2 0.00
            
            0 3 0.00
            1 3 0.00
            2 3 0.00
            3 3 97.20
            4 3 0.00
            5 3 0.00
            6 3 1.87
            7 3 0.93
            8 3 0.00
            9 3 0.00
            
            0 4 0.00
            1 4 2.88
            2 4 0.00
            3 4 0.00
            4 4 97.12
            5 4 0.00
            6 4 0.00
            7 4 0.00
            8 4 0.00
            9 4 0.00

            0 5 0.00
            1 5 0.00
            2 5 0.00
            3 5 0.00
            4 5 3.96
            5 5 96.04
            6 5 0.00
            7 5 0.00
            8 5 0.00
            9 5 0.00

            0 6 0.00
            1 6 0.00
            2 6 0.00
            3 6 0.00
            4 6 0.00
            5 6 0.00
            6 6 100.00
            7 6 0.00
            8 6 0.00
            9 6 0.00

            0 7 0.00
            1 7 0.00
            2 7 4.90
            3 7 2.94
            4 7 0.00
            5 7 0.00
            6 7 1.96
            7 7 90.20
            8 7 0.00
            9 7 0.00

            0 8 2.80
            1 8 0.00
            2 8 0.00
            3 8 0.00
            4 8 0.00
            5 8 0.00
            6 8 0.00
            7 8 0.00
            8 8 97.20
            9 8 0.00

            0 9 0.00
            1 9 2.00
            2 9 0.00
            3 9 2.00
            4 9 1.00
            5 9 0.00
            6 9 0.00
            7 9 0.00
            8 9 0.00
            9 9 95.00

        };
    \end{axis}
\end{tikzpicture}
        \caption{64 GHz with 96.60 \% overall accuracy.}
        \label{conf1}
    \end{subfigure}
    \hfill
    \begin{subfigure}[b]{0.48\linewidth}
        \centering
        \begin{tikzpicture}[scale=0.7]
    \begin{axis}[
            colormap={lightpurple}{color=(white) rgb255=(128,159,106)},
            xlabel=Prediction (in \%),
            xlabel style={yshift=-45pt},
            ylabel=Label,
            ylabel style={yshift=25pt},
            xticklabels={Hammer, Screwdriver, Deodorant, Calculator, Water Bottle, Plastic Cup, Cable, Ball, Mug, Tape},
            xtick={0,...,9},
            xtick style={draw=none},
            yticklabels={Hammer, Screwdriver, Deodorant, Calculator, Water Bottle, Plastic Cup, Cable, Ball, Mug, Tape},
            ytick={0,...,9},
            font =\small,
            ytick style={draw=none},
            enlargelimits=false,
            xticklabel style={
              rotate=90
            },
            nodes near coords={\pgfmathprintnumber\pgfplotspointmeta},
            nodes near coords style={
                yshift=-7pt,
                font=\scriptsize
            },
        ]
        \addplot[
            matrix plot,
            mesh/cols=10,
            point meta=explicit,draw=gray
        ] table [meta=C] {
            x y C
            0 0 88.12
            1 0 0.00
            2 0 0.00
            3 0 0.00
            4 0 0.00
            5 0 0.00
            6 0 0.00
            7 0 9.90
            8 0 1.98
            9 0 0.00
            
            0 1 0.00
            1 1 98.92
            2 1 1.08
            3 1 0.00
            4 1 0.00
            5 1 0.00
            6 1 0.00
            7 1 0.00
            8 1 0.00
            9 1 0.00
            
            0 2 0.00
            1 2 4.04
            2 2 92.93
            3 2 0.00
            4 2 0.00
            5 2 1.01
            6 2 0.00
            7 2 0.00
            8 2 2.02
            9 2 0.00
            
            0 3 0.00
            1 3 0.00
            2 3 0.00
            3 3 94.39
            4 3 1.87
            5 3 0.00
            6 3 0.93
            7 3 0.00
            8 3 0.00
            9 3 2.80
            
            0 4 0.00
            1 4 0.00
            2 4 0.00
            3 4 0.00
            4 4 93.27
            5 4 1.92
            6 4 4.81
            7 4 0.00
            8 4 0.00
            9 4 0.00

            0 5 0.00
            1 5 0.00
            2 5 0.00
            3 5 0.00
            4 5 0.99
            5 5 96.04
            6 5 2.97
            7 5 0.00
            8 5 0.00
            9 5 0.00

            0 6 0.00
            1 6 0.00
            2 6 0.00
            3 6 0.00
            4 6 4.65
            5 6 8.14
            6 6 87.21
            7 6 0.00
            8 6 0.00
            9 6 0.00

            0 7 1.96
            1 7 0.98
            2 7 0.00
            3 7 0.00
            4 7 0.00
            5 7 0.00
            6 7 0.00
            7 7 97.06
            8 7 0.00
            9 7 0.00

            0 8 4.67
            1 8 0.00
            2 8 3.74
            3 8 0.00
            4 8 0.00
            5 8 0.00
            6 8 0.00
            7 8 0.00
            8 8 90.65
            9 8 0.93

            0 9 1.00
            1 9 0.00
            2 9 0.00
            3 9 2.00
            4 9 0.00
            5 9 0.00
            6 9 0.00
            7 9 0.00
            8 9 0.00
            9 9 97.00

        };
    \end{axis}
\end{tikzpicture}
        \caption{67 GHz with 93.59 \% overall accuracy.}
        \label{fig:conf2}
    \end{subfigure}
    \caption{Confusion matrices for ACCOR with weighting factor $\alpha=0.4$.}
    \label{fig:conf-matrices}
\end{figure*}
\begin{table}[t!] 
\centering 
\begin{tabular}{c cc} 
\toprule 
\textbf{Model} & \textbf{64 GHz} & \textbf{67 GHz} \\ 
\midrule 
RadarCNN \cite{radarcnn} & 90.14 \% & 91.80 \%\\ 
SMCNet \cite{smcnet} & 91.89 \% & 92.67 \% \\
Dual-stream CNN \cite{stefan}  & 95.15 \% & 92.30 \% \\
ACCOR (ours) & \textbf{96.60} \% & \textbf{93.59} \% \\
\end{tabular}%
\caption{Overall accuracy of radar models.}
\label{results-1}
\end{table}
\begin{table}[t!] 
\centering 
\begin{tabular}{c cc} 
\toprule 
\textbf{Model} & \textbf{64 GHz} & \textbf{67 GHz} \\ 
\midrule 
Baseline EfficientNet b0 \cite{efficientnet} & 70.83 \% & 55.33 \%\\
Scaled EfficientNet b4 \cite{efficientnet} & 46.83 \% & 36.00 \%\\
ResNet-18 \cite{resnet} & 93.36 \% & 89.00 \% \\ 
ResNet-34 \cite{resnet} & 46.81 \% & 58.67 \% \\ 
ACCOR (ours)& \textbf{96.60} \% & \textbf{93.59} \% \\
\bottomrule 
\end{tabular}%
\caption{Overall accuracy of common image classification models.}
\label{results-2}
\end{table}
Furthermore, Table~\ref{results-2} presents the performance of image classification models conventionally designed for RGB image inputs. To adapt the input and satisfy this requirement, the radar signal is separated into real and imaginary components, which are then zero-padded in the third color channel to match the expected input dimensions.
We deliberately avoided converting the data into an RGB radar map image, since this would discard phase information and reduce the representation to a real-valued image, thereby losing relevant information.
Table~\ref{results-2} presents the results obtained from various image classification models applied to the adapted radar data. ACCOR achieves higher overall accuracy compared to all image classification models. ResNet-18 is the best-performing image classification model on our pseudo-image data, but cannot surpass the ACCOR performance.
Generally, all image-based approaches struggle to extract the features necessary for reliable classification, as expected given the inadequate and ill-suited input representation.
In the end, we also tried to evaluate Transformer models such as Vision Transformers (ViTs) on this task. However, training proved unsuccessful due to the dataset size, as these large architectures require substantially more data to learn generalizable representations and avoid overfitting. A significantly larger dataset would be necessary to properly evaluate ViT performance on mmWave radar occluded object classification.
\subsection{Ablation Study}
\subsubsection{Weighting Factor $\alpha$}
For further evaluation, we conducted an ablation study on the weighting factor $\alpha \in \{0.0, 0.1, \dots, 0.6\}$ in our hybrid loss function (Eq.~\ref{loss}). Setting $\alpha = 0$ removes the contrastive term and corresponds to a pure cross-entropy loss, while $\alpha = 0.5$ gives equal weight to cross-entropy and supervised contrastive loss. 
The results are listed in Table~\ref{results-3}.
Across all settings with $\alpha > 0$, both frequency bands outperform the cross-entropy-only baseline, which confirms that the contrastive term systematically improves feature discriminability for radar data. The 64 GHz subset achieves slightly higher accuracy than the 67 GHz subset for most $\alpha$'s, indicating that the model can exploit the signal characteristics at 64 GHz more effectively.
Performance peaks for intermediate weights, with best results around $\alpha = 0.3$$–$$0.5$ for 64 GHz. The maximum accuracy of 96.60\% is obtained at $\alpha = 0.4$, while for 67 GHz the best value of 93.89\% occurs at $\alpha = 0.5$. When $\alpha$ is chosen too small ($\leq0.1$) or too large ($\geq0.6$), accuracy drops for both frequencies, suggesting that relying exclusively on one of the two loss terms either under-utilizes the contrastive signal or destabilizes optimization. The small discrepancy between the optimal $\alpha$ for 64 GHz and 67 GHz (0.4 vs. 0.5) is below 0.5 percentage points and thus within typical training variability. We therefore interpret both as indicating an optimal weighting in which cross-entropy and contrastive objectives contribute with comparable strength. Overall, we set $\alpha = 0.4$ as the best trade-off that yields consistently strong performance on both frequency bands.
\begin{table}[t!] 
\centering 
\begin{tabular}{c cc} 
\toprule 
\textbf{ACCOR Loss Weighting} & \textbf{64 GHz} & \textbf{67 GHz} \\ 
\midrule 
$\alpha = 0.6$ &  96.10 \% &  92.89 \% \\
$\alpha = 0.5$ &  96.17 \% &  93.89 \% \\
$\alpha = 0.4$  & 96.60 \% & 93.59 \% \\
$\alpha = 0.3$  & 96.40 \% & 93.39 \% \\
$\alpha = 0.2$ & 96.40 \% & 92.29 \% \\
$\alpha = 0.1$ & 95.30 \% &  91.60 \%\\  
$\alpha = 0$ (only CE) & 94.50 \% & 90.10 \% \\
\bottomrule 
\end{tabular}%
\caption{Ablation study of our proposed method for different values of $\alpha$.}
\label{results-3}
\end{table}
\begin{figure*}
    \centering
    \begin{subfigure}[b]{0.45\linewidth}
        \centering
        \includegraphics[width=\linewidth]{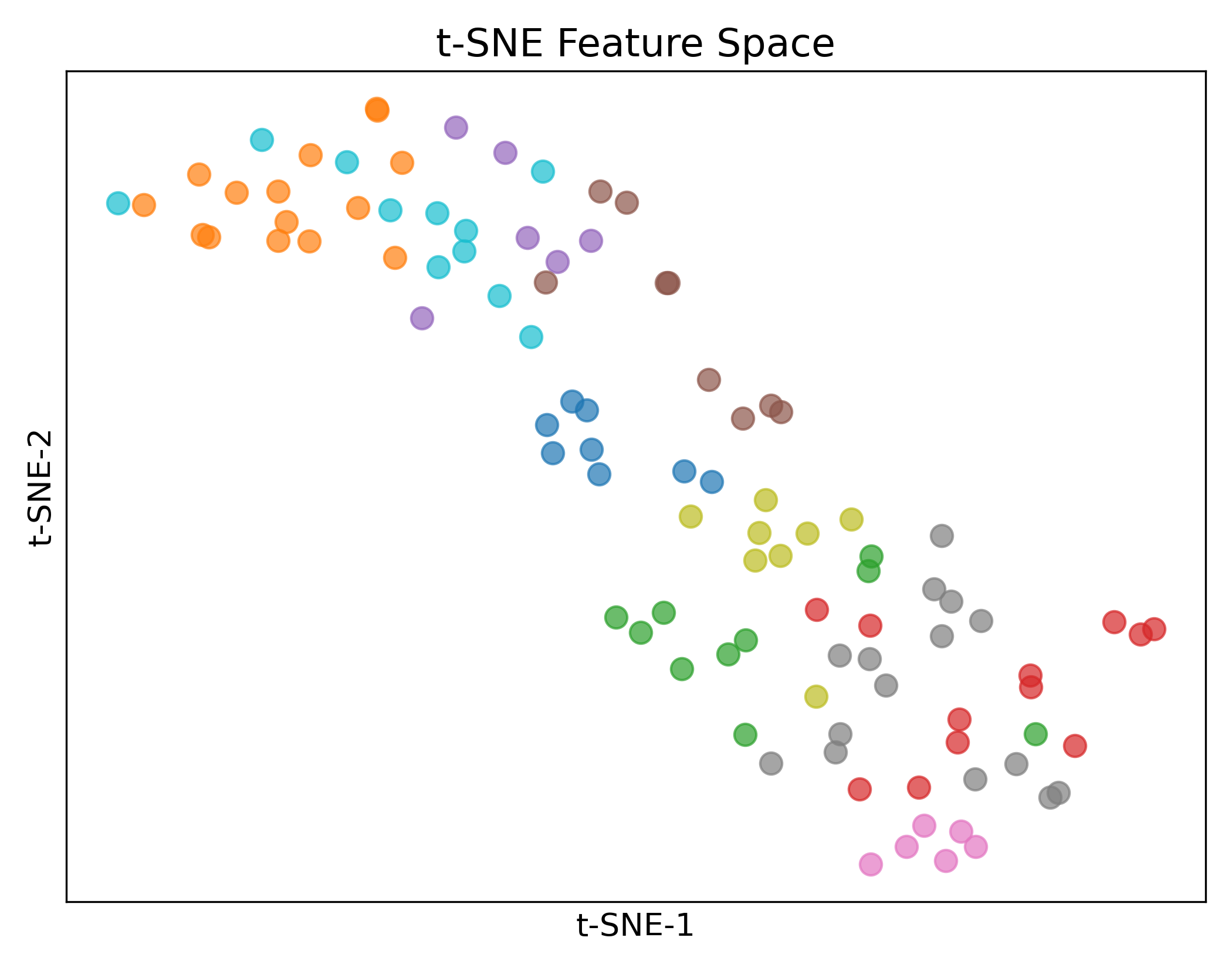}
        \caption{$\alpha=0$}
        \label{fig:sub1}
    \end{subfigure}
    \hfill
    \begin{subfigure}[b]{0.45\linewidth}
        \centering
        \includegraphics[width=\linewidth]{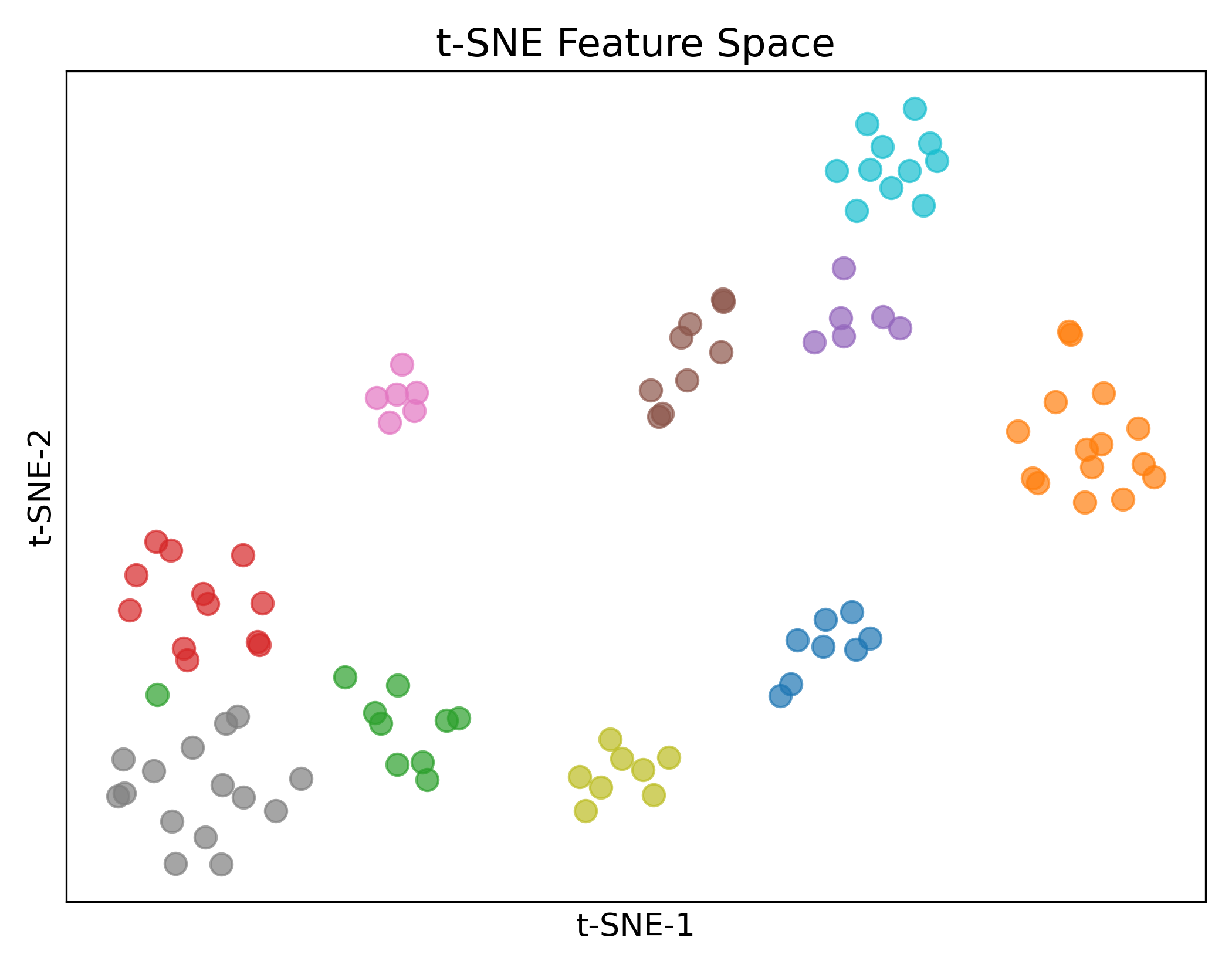}
        \caption{$\alpha=0.4$}
        \label{fig:sub2}
    \end{subfigure}
    \caption{Comparison of the t-SNE feature space graph for all 10 classes at 64 GHz with perplexity 20 and after 2000 iterations.}
    \label{tsne-plots}
\end{figure*}
To visualize the impact of the contrastive term on the model’s feature space, we project the learned features onto a 2D t-SNE graph with perplexity 20 and 2000 iterations. The graphs are depicted in Fig.~\ref{tsne-plots}.
Here, a small Euclidean distance between points indicates high similarity in the learned feature space. Each color represents a class from the dataset.
The feature space trained with contrastive loss ($\alpha=0.4$) shows clearly superior clustering and reduced distances between same-class samples compared to training with CE loss only ($\alpha=0$).
A greater separation between classes is clearly visible.
This visualization highlights the effectiveness of the contrastive learning approach.
\subsubsection{Real-Valued Backbone}
To further justify the use of a complex-valued CNN as the model backbone, we additionally compare our approach against a purely real-valued backbone counterpart using the same architecture.
The design is identical to the complex-valued version in terms of number of parameters and kernel sizes. The primary difference lies in the fact that all operations are performed in the real domain, with the input provided as a two-channel tensor comprising real- and imaginary parts.
Moreover, we use the weighting factor $\alpha=0.4$ and $\alpha=0.5$ for comparison, which yielded the best results using the complex-valued backbone.
\begin{table}[t!] 
\centering 
\resizebox{\columnwidth}{!}{%
\begin{tabular}{c cc} 
\toprule 
\textbf{Model} & \textbf{64 GHz} & \textbf{67 GHz} \\ 
\midrule 
ACCOR (real-valued backbone, $\alpha=0.4$)  & 90.70 \% & 91.29 \%\\
ACCOR (real-valued backbone, $\alpha=0.5$)  & 90.00 \% & 91.40 \%\\
\bottomrule 
\end{tabular}%
}
\caption{Ablation study using a real-valued CNN backbone with adjusted input tensor and same hybrid loss $\ell_{\text{total}}$.}
\label{results-4}
\end{table}
Table~\ref{results-4} provides the accuracy for the two frequency bands. Keeping the rest of the architecture identical, performance drops substantially when switching from a complex-valued to a real-valued feature extractor. This indicates a substantially improved feature extraction capability by the complex-valued network, which aligns well with the inherently complex-valued nature of the input signals.
\subsubsection{Remarks}
While ACCOR shows strong performance and complex-valued processing capability, the modest dataset size and single cardboard type reflect typical proof-of-concept constraints. The narrow 64–67 GHz spacing (0.2 mm wavelength gap) yields subtle penetration differences, demonstrating fundamental viability even within narrow bands. These motivate future expansion of data and model while establishing a solid foundation with this work.

\section{Conclusion}
\label{conclusio}
In this work, we addressed occluded object classification for non-visual inspection in indoor settings, demonstrating that mmWave radar is a suitable sensor for this task. Our proposed ACCOR model, which integrates complex-valued CNNs, attention, and a hybrid loss, achieved an accuracy of 96.60 \% at 64 GHz and 93.59 \% at 67 GHz, surpassing existing radar and image-based classifiers. While the small dataset size and $\alpha$ ambiguity highlight opportunities for further refinement, and the limited frequency range calls for broader evaluation, these results confirm the promise of complex-valued attention-enhanced contrastive learning for mmWave radar perception. Future work will expand to collecting larger datasets with more significant wavelength differences and more diverse objects and different occlusion types, leading the way toward robust radar-based perception for different environments.

\section*{Acknowledgement}
\thanks{The authors would like to thank the Federal Ministry of Research, Technology, and Space (BMFTR) for its support as part of the research program Communication Systems “Souverän. Digital. Vernetzt.”. Joint project 6G-life, project identification number: 16KIS2414}
{
    \small
    \bibliographystyle{ieeenat_fullname}  %
    \bibliography{main.bib}

@String(CVPR= {IEEE Conf. Comput. Vis. Pattern Recog.})

@String(ICASSP=	{ICASSP})

@String(CVPR  = {CVPR})

@inproceedings{human,
  author    = {A. -C. Froehlich and others},
  title     = {A Millimeter-Wave {MIMO} Radar Network for Human Activity Recognition and Fall Detection},
  booktitle = {2024 {IEEE} Radar Conference (RadarConf24)},
  year      = {2024},
  pages     = {1--5},
  doi       = {10.1109/RadarConf2458775.2024.10548702}
}

@inproceedings{tracking,
  author    = {P. Zhao and others},
  title     = {{mID}: Tracking and Identifying People with Millimeter Wave Radar},
  booktitle = {2019 15th International Conference on Distributed Computing in Sensor Systems (DCOSS)},
  year      = {2019},
  pages     = {33--40},
  doi       = {10.1109/DCOSS.2019.00028}
}

@inproceedings{tracking2,
  author    = {Z. Pan and F. Ding and H. Zhong and C. X. Lu},
  title     = {{RaTrack}: Moving Object Detection and Tracking with {4D} Radar Point Cloud},
  booktitle = {2024 IEEE International Conference on Robotics and Automation (ICRA)},
  year      = {2024},
  pages     = {4480--4487},
  doi       = {10.1109/ICRA57147.2024.10610368}
}

@inproceedings{gesture1,
  author    = {L. Senigagliesi and G. Ciattaglia and D. Disha and E. Gambi},
  title     = {Classification of Human Activities based on Automotive Radar Spectral Images Using Machine Learning Techniques: A Case Study},
  booktitle = {2022 {IEEE} Radar Conference (RadarConf22)},
  year      = {2022},
  pages     = {1--6},
  doi       = {10.1109/RadarConf2248738.2022.9764217}
}

@inproceedings{gesture2,
  author    = {J. Chen and P. Wen and G. Chen and Y. Wang and Y. Wang and J. Zheng},
  title     = {Hand Gesture Recognition based on Millimeter-Wave Radar using {iFormer}},
  booktitle = {2024 9th International Conference on Signal and Image Processing (ICSIP)},
  year      = {2024},
  pages     = {22--26},
  doi       = {10.1109/ICSIP61881.2024.10671474}
}

@inproceedings{gesture3,
  author    = {S. M. Kahya and M. Sami Yavuz and E. Steinbach},
  title     = {{HAROOD}: Human Activity Classification and Out-Of-Distribution Detection with Short-Range {FMCW} Radar},
  booktitle = {2024 IEEE International Conference on Acoustics, Speech and Signal Processing (ICASSP)},
  year      = {2024},
  pages     = {6950--6954},
  doi       = {10.1109/ICASSP48485.2024.10447729}
}

@inproceedings{philipp,
  author    = {P. Wolters and others},
  title     = {Unleashing {HyDRa}: Hybrid Fusion, Depth Consistency and Radar for Unified 3D Perception},
  booktitle = {2025 {IEEE} International Conference on Robotics and Automation (ICRA)},
  year      = {2025},
  pages     = {7467--7474},
  doi       = {10.1109/ICRA55743.2025.11127412}
}

@INPROCEEDINGS{philipp2,
  author={P. Wolters and J. Gilg and T. Teepe and G. Rigoll},
  booktitle={2025 IEEE/CVF International Conference on Computer Vision Workshops (ICCVW)}, 
  title={{SpaRC-AD}: A Baseline for Radar-Camera Fusion in End-to-End Autonomous Driving}, 
  year={2025},
  volume={},
  number={},
  pages={1831-1841},
  doi={10.1109/ICCVW69036.2025.00193}}

@misc{survey,
  author       = {X. Peng and M. Tang and H. Sun and L. Servadei and R. Wille},
  title        = {{4D} mmWave Radar in Adverse Environments for Autonomous Driving: A Survey},
  year         = {2025},
  note         = {arXiv preprint}
}

@inproceedings{mengchen,
  author    = {M. Xiong and X. Xu and D. Yang and E. Steinbach},
  title     = {Robust Depth Estimation in Foggy Environments Combining {RGB} Images and mmWave Radar},
  booktitle = {2022 {IEEE} International Symposium on Multimedia (ISM)},
  year      = {2022},
  pages     = {34--41},
  doi       = {10.1109/ISM55400.2022.00011}
}

@inproceedings{smoke,
  author    = {H. Kulhandjian and A. Davis and L. Leong and M. Bendot and M. Kulhandjian},
  title     = {{AI}-based Human Detection and Localization in Heavy Smoke using Radar and {IR} Camera},
  booktitle = {2023 {IEEE} Radar Conference (RadarConf23)},
  year      = {2023},
  pages     = {1--6},
  doi       = {10.1109/RadarConf2351548.2023.10149735}
}

@inproceedings{smoke2,
  author    = {X. Shuai and Y. Shen and Y. Tang and S. Shi and L. Ji and G. Xing},
  title     = {{MilliEye}: A Lightweight mmWave Radar and Camera Fusion System for Robust Object Detection},
  booktitle = {Proceedings of the International Conference on Internet-of-Things Design and Implementation (IoTDI)},
  year      = {2021},
  pages     = {145--157},
  doi       = {10.1145/3450268.3453532}
}

@article{imaging1,
  author  = {R. Rückert and I. Ullmann and C. Herglotz and A. Kaup and M. Vossiek},
  title   = {Data Compression for Close-Range Radar Imaging},
  journal = {{IEEE} Transactions on Radar Systems},
  year    = {2024},
  volume  = {2},
  pages   = {421--433},
  doi     = {10.1109/TRS.2024.3387288}
}

@article{imaging2,
  author  = {S. S. Ahmed},
  title   = {Microwave Imaging in Security {-} Two Decades of Innovation},
  journal = {{IEEE} Journal of Microwaves},
  year    = {2021},
  volume  = {1},
  number  = {1},
  pages   = {191--201},
  doi     = {10.1109/JMW.2020.3035790}
}

@article{stefan,
  author  = {S. Hägele and F. Seguel and S. M. Kahya and E. Steinbach},
  title   = {Occluded Object Classification With mmWave {MIMO} Radar {IQ} Signals Using Dual-Stream Convolutional Neural Networks},
  journal = {{IEEE} Transactions on Radar Systems},
  year    = {2025},
  volume  = {3},
  pages   = {789--798},
  doi     = {10.1109/TRS.2025.3571284}
}

@inproceedings{cplx_cnn,
  author    = {F. Seguel and D. Salihu and S. Haegele and E. Steinbach},
  title     = {Complex-valued Deep Learning for {WiFi}-based Indoor Positioning: Method and Performance},
  booktitle = {29th European Wireless Conference (European Wireless 2024)},
  year      = {2024},
  pages     = {59--65}
}

@inproceedings{cplx_cnn2,
  author    = {S. Chatterjee and P. Tummala and O. Speck and A. Nürnberger},
  title     = {Complex Network for Complex Problems: A Comparative Study of {CNN} and Complex-valued {CNN}},
  booktitle = {2022 {IEEE} 5th International Conference on Image Processing Applications and Systems (IPAS)},
  year      = {2022},
  pages     = {1--5},
  doi       = {10.1109/IPAS55744.2022.10053060}
}

@inproceedings{smcnet,
  author    = {S. Hägele and F. Seguel and D. Salihu and A. Misik and E. Steinbach},
  title     = {{SMCNet}: Supervised Surface Material Classification Using mmWave Radar {IQ} Signals and Complex-valued {CNN}s},
  booktitle = {2025 {IEEE} International Conference on Acoustics, Speech and Signal Processing (ICASSP)},
  year      = {2025},
  pages     = {1--5},
  doi       = {10.1109/ICASSP49660.2025.10890769}
}

@article{fusion1,
  author  = {Y. Liu and S. Chang and Z. Wei and K. Zhang and Z. Feng},
  title   = {Fusing mmWave Radar With Camera for {3-D} Detection in Autonomous Driving},
  journal = {{IEEE} Internet of Things Journal},
  year    = {2022},
  volume  = {9},
  number  = {20},
  pages   = {20408--20421},
  doi     = {10.1109/JIOT.2022.3175375}
}

@inproceedings{weather,
  author    = {T. Kawaguchi and K. Shinotsuka and S. Malterer},
  title     = {Experimental Verification of Rainfall Impact on Sparse Array Radar},
  booktitle = {2024 {IEEE} Radar Conference (RadarConf24)},
  year      = {2024},
  pages     = {1--6},
  doi       = {10.1109/RadarConf2458775.2024.10548312}
}

@article{tracking3,
  author  = {J. Pegoraro and M. Rossi},
  title   = {Real-Time People Tracking and Identification From Sparse mm-Wave Radar Point-Clouds},
  journal = {{IEEE} Access},
  year    = {2021},
  volume  = {9},
  pages   = {78504--78520},
  doi     = {10.1109/ACCESS.2021.3083980}
}

@article{gesture4,
  author  = {S. Ahmed and S. Abdullah and S. H. Cho},
  title   = {Advancements in Radar Point Cloud Processing for Macro Human Movements in Healthcare and Assisted Living Domains: A Review},
  journal = {{IEEE} Sensors Journal},
  year    = {2024},
  volume  = {24},
  number  = {22},
  pages   = {36287--36305},
  doi     = {10.1109/JSEN.2024.3452110}
}

@inproceedings{gesture5,
  author    = {G. Liao and J. Ma and F. Luo},
  title     = {Human Activity Recognition by Using Enhanced Radar Point Cloud {2D} Histograms and Doppler Feature Fusion},
  booktitle = {2025 {IEEE} International Conference on Robotics and Automation (ICRA)},
  year      = {2025},
  pages     = {11234--11241},
  doi       = {10.1109/ICRA55743.2025.11128165}
}

@inproceedings{sar1,
  author    = {D. Qosja and S. Wagner and D. O'Hagan},
  title     = {{SAR} Image Synthesis with Diffusion Models},
  booktitle = {2024 {IEEE} Radar Conference (RadarConf24)},
  year      = {2024},
  pages     = {1--6},
  doi       = {10.1109/RadarConf2458775.2024.10549257}
}

@inproceedings{sar2,
  author    = {X. Wang and T. Ye and R. Kannan and V. Prasanna},
  title     = {{FACTUAL}: A Novel Framework for Contrastive Learning Based Robust {SAR} Image Classification},
  booktitle = {2024 {IEEE} Radar Conference (RadarConf24)},
  year      = {2024},
  pages     = {1--6},
  doi       = {10.1109/RadarConf2458775.2024.10549364}
}

@article{imaging3,
  author  = {W. Zhang and Y. Ji and W. Shao and B. Lin and C. Li and G. Fang},
  title   = {A Fast {3-D} Chirp Scaling Imaging Technique for Millimeter-Wave Near-Field Imaging},
  journal = {{IEEE} Transactions on Microwave Theory and Techniques},
  year    = {2023},
  volume  = {71},
  number  = {2},
  pages   = {827--841},
  doi     = {10.1109/TMTT.2022.3205926}
}

@article{imaging4,
  author  = {F. Zhang and C. Wu and B. Wang and K. J. R. Liu},
  title   = {{mmEye}: Super-Resolution Millimeter Wave Imaging},
  journal = {{IEEE} Internet of Things Journal},
  year    = {2021},
  volume  = {8},
  number  = {8},
  pages   = {6995--7008},
  doi     = {10.1109/JIOT.2020.3037836}
}

@misc{vayyar,
  author = {{Mini-Circuits} and {Vayyar}},
  title  = {{{IMAGEVK-74} 4D Imaging Radar}},
  note   = {Available online: \url{https://www.minicircuits.com/WebStore/imagevk_74.html}}
}

@inproceedings{radarcnn,
  author    = {S. Hägele and F. Seguel and D. Salihu and M. Zakour and E. Steinbach},
  title     = {{RadarCNN}: Learning-Based Indoor Object Classification from {IQ} Imaging Radar Data},
  booktitle = {2024 {IEEE} Radar Conference (RadarConf24)},
  year      = {2024},
  pages     = {1--6},
  doi       = {10.1109/RadarConf2458775.2024.10548874}
}

@inproceedings{iq,
  author    = {Z. Huang and A. Pemasiri and S. Denman and C. Fookes and T. Martin},
  title     = {Multi-Task Learning for Radar Signal Characterisation},
  booktitle = {2023 {IEEE} International Conference on Acoustics, Speech, and Signal Processing Workshops (ICASSPW)},
  year      = {2023},
  pages     = {1--5},
  doi       = {10.1109/ICASSPW59220.2023.10193318}
}

@article{wall1,
  author  = {M. A. Maisto and M. Masoodi and R. Pierri and R. Solimene},
  title   = {Sensor Arrangement in Through-the-Wall Radar Imaging},
  journal = {{IEEE} Open Journal of Antennas and Propagation},
  year    = {2022},
  volume  = {3},
  pages   = {333--341},
  doi     = {10.1109/OJAP.2022.3159279}
}

@inproceedings{wall2,
  author    = {S. Huang and J. Qian and Y. Wang and X. Yang and L. Yang},
  title     = {Through-the-Wall Radar Super-Resolution Imaging Based on Cascade {U-Net}},
  booktitle = {2019 {IEEE} International Geoscience and Remote Sensing Symposium (IGARSS)},
  year      = {2019},
  pages     = {2933--2936},
  doi       = {10.1109/IGARSS.2019.8900569}
}

@article{tro_radar,
  author  = {K. Harlow and H. Jang and T. D. Barfoot and A. Kim and C. Heckman},
  title   = {A New Wave in Robotics: Survey on Recent {mmWave} Radar Applications in Robotics},
  journal = {{IEEE} Transactions on Robotics},
  year    = {2024},
  volume  = {40},
  pages   = {4544--4560},
  doi     = {10.1109/TRO.2024.3463504}
}

@inproceedings{robots,
  author    = {X. Huang and N. Patel and K. P. Tsoi},
  title     = {Application of mmWave Radar Sensors for Autonomous Robotics Navigation},
  booktitle = {2023 {IEEE} Asia-Pacific Conference on Computer Science and Data Engineering (CSDE)},
  year      = {2023},
  pages     = {1--7},
  doi       = {10.1109/CSDE59766.2023.10487689}
}

@inproceedings{drones,
  author = {P. Meiresone and D. Van Hamme and W. Philips and T. Verbelen},
  title = {Ego-Motion Estimation with a Low-Power Millimeter-Wave Radar on a {UAV}},
  booktitle = {International Conference on Radar Systems (RADAR 2022)},
  year = {2022},
  pages = {371--376},
  doi = {10.1049/icp.2022.2346}
}

@article{cont1,
  author  = {Y. Li and C. Wan and X. Zhou and T. Tang},
  title   = {Small-Sample {SAR} Target Recognition Using a Multimodal Views Contrastive Learning Method},
  journal = {{IEEE} Geoscience and Remote Sensing Letters},
  year    = {2025},
  volume  = {22},
  pages   = {1--5},
  note    = {Art. no. 4007905},
  doi     = {10.1109/LGRS.2025.3557534}
}

@article{cont2,
  author  = {C. Li and L. Du and Y. Du},
  title   = {Semi-Supervised {SAR} {ATR} Based on Contrastive Learning and Complementary Label Learning},
  journal = {{IEEE} Geoscience and Remote Sensing Letters},
  year    = {2024},
  volume  = {21},
  pages   = {1--5},
  note    = {Art. no. 4016405},
  doi     = {10.1109/LGRS.2024.3458948}
}

@article{transrad,
  author  = {L. Cheng and S. Cao},
  title   = {{TransRAD}: Retentive Vision Transformer for Enhanced Radar Object Detection},
  journal = {{IEEE} Transactions on Radar Systems},
  year    = {2025},
  volume  = {3},
  pages   = {303--317},
  doi     = {10.1109/TRS.2025.3537604}
}

@inproceedings{attention,
  author    = {A. Vaswani and N. Shazeer and N. Parmar and J. Uszkoreit and L. Jones and A. N. Gomez and Ł. Kaiser and I. Polosukhin},
  title     = {Attention Is All You Need},
  booktitle = {Advances in Neural Information Processing Systems (NeurIPS)},
  year      = {2017}
}

@inproceedings{resnet,
  author    = {K. He and X. Zhang and S. Ren and J. Sun},
  title     = {Deep Residual Learning for Image Recognition},
  booktitle = {2016 {IEEE} Conference on Computer Vision and Pattern Recognition (CVPR)},
  year      = {2016},
  pages     = {770--778},
  doi       = {10.1109/CVPR.2016.90}
}

@misc{efficientnet,
  author       = {M. Tan and Q. V. Le},
  title        = {{EfficientNet}: Rethinking Model Scaling for Convolutional Neural Networks},
  year         = {2019},
  archivePrefix = {arXiv},
  eprint       = {1905.11946}
}
}

\end{document}